\documentclass[12pt,english,aps,prb,preprint,superscriptaddress,showpacs,showkeys]{article}
\usepackage[LGR,T1]{fontenc}
\usepackage[latin9]{inputenc}
\usepackage{amsmath}
\usepackage{amssymb}
\usepackage{graphicx}

\makeatletter

\DeclareRobustCommand{\greektext}{%
  \fontencoding{LGR}\selectfont\def\encodingdefault{LGR}}
\DeclareRobustCommand{\textgreek}[1]{\leavevmode{\greektext #1}}
\DeclareFontEncoding{LGR}{}{}
\DeclareTextSymbol{\~}{LGR}{126}

\newcommand{\lyxaddress}[1]{
\par {\raggedright #1
\vspace{1.4em}
\noindent\par}
}

\makeatletter

\usepackage{amsfonts}

\usepackage{times}
\usepackage{babel}
\usepackage[super,comma,sort&compress]{natbib}

\usepackage{epstopdf}

\makeatletter

 \usepackage{epstopdf}
    \usepackage{times}
\makeatother

\makeatother

\usepackage{babel}
\begin{document}

\title{Predicting phase behavior of grain boundaries with evolutionary search
and machine learning}

\author{Qiang Zhu$^{1}$,\enskip{}Amit Samanta$^{2}$,\enskip{}Bingxi Li$^{3}$,
Robert E. Rudd$^{2}$ and Timofey Frolov$^{2}$}

\maketitle

\lyxaddress{$^{1}$Department of Physics and Astronomy, High Pressure Science
and Engineering Center, University of Nevada, Las Vegas, NV 89154,
USA}

\lyxaddress{$^{2}$ Lawrence Livermore National Laboratory, Livermore, California
94550, USA}

\lyxaddress{$^{3}$ Department of Computer Science, University of California
Davis, Davis, CA 95616, USA}

\bigskip{}

\noindent \begin{center}
\textsf{\textbf{Abstract}}
\par\end{center}

The study of grain boundary phase transitions is an emerging field
until recently dominated by experiments. The major bottleneck in exploration
of this phenomenon with atomistic modeling has been the lack of a
robust computational tool that can predict interface structure. Here
we develop a new computational tool based on evolutionary algorithms
that performs efficient grand-canonical grain boundary structure search
and we design a clustering analysis that automatically identifies
different grain boundary phases. Its application to a model system
of symmetric tilt boundaries in Cu uncovers an unexpected rich polymorphism
in the grain boundary structures. We find new ground and metastable
states by exploring structures with different atomic densities. Our
results demonstrate that the grain boundaries within the entire misorientation
range have multiple phases and exhibit structural transitions, suggesting
that phase behavior of interfaces is likely a general phenomenon.

\newpage{}

\noindent \begin{flushleft}
\textsf{\textbf{\large{}Introduction}}\textsf{\textbf{ }}
\par\end{flushleft}

The pursuit of new technologies for cleaner energy conversion and
more efficient energy utilization has generated increased interest
in the development of advanced metallic alloys and ceramics that can
operate safely at high temperatures and in aggressive environments.
The properties of these structural and functional materials are strongly
influenced by the presence of internal interfaces called grain boundaries,
which are inherited from materials synthesis and processing. Understanding
the structure of these interfaces and the ways it influences properties
can be key to optimizing materials to meet the needs of advanced energy
applications.

Recent years have seen a rapid growth of evidence suggesting that
grain boundaries can exist in multiple states or phases and exhibit
first-order transitions, marked by discontinuous changes in properties
like segregation, mobility, cohesive strength and sliding resistance\cite{Cantwell20141}.
These discontinuous transitions were observed in isolated bicrystals
with a single well-defined grain boundary as well as in polycrystalline
samples with many different grain boundaries. For example, measurements
of Ag impurity diffusion in the $\Sigma5(310)[001]$ grain boundary
(GB) in Cu revealed an unusual non-Arrhenius behavior of the diffusion
flux characterized by two distinct slopes at low and high temperatures\cite{Divinski2012}.
In polycrystals, studies of doped ceramics demonstrated non-Arrhenius
behavior of growth rate constant which exhibits multiple discontinuous
transitions with temperature\cite{Dillon20075247,Dillon20076208}.
High-resolution transmission electron microscopy (HRTEM) analysis
of these ceramics identified GB structures resembling intergranular
films of different thickness\cite{Cantwell20141}. The discontinuous
nature of these transitions in polycrystalline materials is somewhat
unexpected. If the changes in the grain growth behavior were indeed
triggered by transformations of the interface structure, one would
expect more gradual changes in properties, since at different interfaces
in the material the transitions should take place at different temperatures
and impurity concentrations. The discontinuous character of the mobility
jumps measured in the experiments on the other hand suggests that
the transitions at different interfaces may happen in a more uniform
manner.

To explain this puzzling behavior it was proposed that grain boundaries
can exist in multiple states called \textit{complexions}\cite{Harmer08042011,Dillon20075247,Dillon20076208,Cantwell20141}.
Complexion types are characterized by different amounts of impurity
segregation. Monolayer, bilayer, trilayer and thicker films types
of complexions have been suggested\cite{Dillon20076208}. Grain boundary
complexions were predicted by earlier theoretical work. Phase field
models have led to predictions of a variety of first-order and higher
order premelting type transitions and mapped them onto bulk phase
diagrams\cite{Tang06,Tang06b}. More recently, layering transitions
associated with GB segregation were investigated using lattice gas
models\cite{Rickman201388,Rickman20161,Rickman2016225} and first-principles
calculations\cite{PhysRevB.90.144102}. Transitions between complexions
of different type could be responsible for changes by orders of magnitude
in the grain growth constant with doping. Experimental studies suggested
a potential role of complexions transitions on abnormal grain growth
in ceramics\cite{Dillon20076208}, activated sintering\cite{Luo99},
and liquid metal embrittlement\cite{Luo23092011}. More recently the
notion of GB complexions has been extended to lattice dislocations,
pointing out that they can also exist in multiple states called linear
complexions. These studies suggested a potential importance of these
transition to mechanical properties of materials\cite{Kaplan1059,Kuzmina1080}.

The body of experimental work currently available on grain boundary
phase transitions has raised a number of fundamental questions concerning
the atomistic structure of the different phases, the kinetics of the
transitions, and the ways in which these interfacial processes influence
grain boundary mobilities, diffusivities and mechanical strength.
While the experimental investigation of the role of grain boundary
phase transitions on materials properties is currently a highly active
field of research in the area of structural and functional materials\cite{Baram08042011,Luo23092011,Harmer08042011,Rheinheimer201568,Cantwell20141,Kuzmina1080,Dillon2016324,Rohrer2016231},
the atomic structure of these grain boundary phases remains unknown.
Direct experimental observations of interfacial phase transitions
at high temperature by HRTEM are extremely difficult due to inherent
limitations\cite{PhysRevLett.59.2887}. A large number of HRTEM studies
of grain boundaries in doped metallic and ceramic materials demonstrated
grain boundary structures resembling inter-granular films of different
thickness\cite{Baram08042011,JACE:JACE603,Luo23092011,Cantwell20141,Dillon20076208}.
Unfortunately these HRTEM images often do not provide sufficient information
about the atomic level structure of these boundaries, so it is still
debated whether these grain boundaries are ordered, partially ordered,
amorphous or liquid.

On the other hand, atomistic simulations can be used to predict atomic
structure of interfaces and study their thermodynamic and kinetic
properties. The common approach to construct grain boundaries in atomistic
simulations is called $\gamma$-surface method. It has been employed
to study interfaces in a variety of materials for more than four decades.
The details of this approach will be described later in the article.
A growing number of recent studies proposed alternative approaches
of grain boundary construction demonstrating that the $\gamma$-surface
method is often not sufficient to predict true ground states\cite{Phillpot1992,Phillpot1994,Alfthan06,Chua:2010uq}.

Despite the decades of extensive modeling research, until recently
atomistic simulations did not provide much evidence of first-order
grain boundary phase transitions\cite{Olmsted2011}. Recently, the
investigation of two high-angle boundaries $\Sigma5(210)[001]$ and
$\Sigma5(310)[001]$ in Cu demonstrated that the critical impediment
to observe such transformations was rooted in inadequate simulation
methodology that uses constant number of atoms and periodic boundary
conditions. High-temperature anneals of these boundaries connected
to open surfaces allowed the number of atoms in the grain boundary
to vary by diffusion, achieving lower free energy states. The simulations
revealed multiple new grain boundary phases of the boundaries characterized
by different atomic densities and demonstrated fully reversible first-order
transitions induced by temperature, changes in chemical compositions
and point defects\cite{Frolov2013,PhysRevB.92.020103,Frolov2016}.
This ingenious modeling approach demonstrated phase behavior of two
special high-angle boundaries that have been extensively investigated
in the past, suggesting that the entire phenomenon could have been
overlooked by modeling due an overly restrictive simulation methodology.
This work identified the limitations of current modeling capabilities
and demonstrated that the greatest obstacle to observing grain boundary
phase transitions in simulations is not their absence in the model
systems, but the lack of a robust computational tool that can predict
complex grain boundary structures. 

In the recent years, there have been significant advances in predicting
the structures from first-principles\cite{Reilly:gp5080}. Among them,
our approach based on the evolutionary algorithm USPEX has proved
to be extremely powerful in different systems including bulk crystals\cite{doi:10.1063/1.2210932}
, 2D crystals\cite{Zhou-PRL-2014}, surfaces\cite{Zhu-PRB-2013},
polymers\cite{Zhu-JCP-2014} and clusters\cite{Lyakhov-CPC-2013},
etc. Extending the method to grain boundaries is logically the next
step. There have been a few pioneering works reported in the literature\cite{Chua:2010uq,Alfthan06,PhysRevB.80.174102}.
For instance, Chua et al has developed a genetic algorithm to study
the non-stoichiometric grain boundaries of SrTiO$_{3}$~\cite{Chua:2010uq}.
However, it was only designed for a system with a fixed number of
atoms and supercell size. In this work, we extend it into a more general
way to enable the automated exploration in higher dimensional space,
which includes the structures with variable number of atoms and variable
cell sizes. It is well known that the complexity exponentially increases
with the growing dimensionality\cite{doi:10.1021/ar1001318}. In that
case, a key to ensure efficient sampling is to find balance between
individual quality and population diversity. Any pure random structure
initialization or variation operation is very likely to lead to disordered
like structures with close energetics. To address this challenge,
we followed the idea of coarse-grained modeling and define the simplified
representations during the stage of structure generation. Some key
representations used here are symmetry, vibrational modes and degree
of local order (see Supplementary Note 1 for details)\cite{Zhu-PRB-2015}.
Wrapping up all these ingredients, we developed a new powerful computational
tool based on evolutionary algorithms that predicts structures of
interfaces. This tool generates a population of grain boundary structures
and improves them over several generations to predict low-energy configurations.
During the evolution complex and diverse structures with different
atomic densities are sampled by operations of heredity and mutation
which involve atomic rearrangements as well as addition and removal
of atoms from the grain boundary core. 

\bigskip{}

\noindent \begin{flushleft}
\textsf{\textbf{\large{}Results}}\textsf{\textbf{ }}
\par\end{flushleft}

\noindent \textsf{\textbf{Grain boundary structure calculations. }}We
demonstrate the robustness and the predictive power of this new computational
method by performing a grand-canonical grain boundary structure search
for high-angle and low-angle boundaries within the entire misorientation
range for {[}001{]} symmetric tilt boundaries in Cu modeled with an
embedded-atom (EAM) potential\cite{Mishin01}. This choice of the
model system is motivated by discontinuous changes in properties in
$\Sigma5(310)[001]$ Cu grain boundary measured experimentally\cite{Divinski2012}
and the discovery of multiple phases of this boundary by high-temperature
molecular dynamics (MD) simulations\cite{Frolov2013PRL,Frolov2013}.
The study raised new questions concerning whether these transitions
are characteristic of only high-angle special boundaries with low
$\Sigma$ or a more general phenomena. It is also not clear how the
crystallographic degrees of freedom such as misorientation angle affect
the multiplicity of grain boundary phases and their properties. With
the newly developed tool, we aim to identify possible multiple grain
boundary phases and recover grain boundary energy as a function of
misorientation as well as atomic density, which have been completely
ignored by the conventional methodology. 

To make a comparison and illustrate the potential importance of this
advanced sampling, we first present the results when the grain boundaries
are constructed using the common methodology. In this approach often
referred to as the $\gamma$-surface method, the two misoriented crystals
are joined together, while sampling relative translations of the grains.
The prepared configurations composed of two grains are then statically
relaxed. During the relaxation the atoms in the boundary fall into
the local minima, which concludes the construction. During the search
no atoms are added or removed from the grain boundary core. Thus,
the grain boundary structures with different atomic densities are
not sampled.

\begin{figure}
\begin{centering}
\includegraphics[width=0.8\paperwidth]{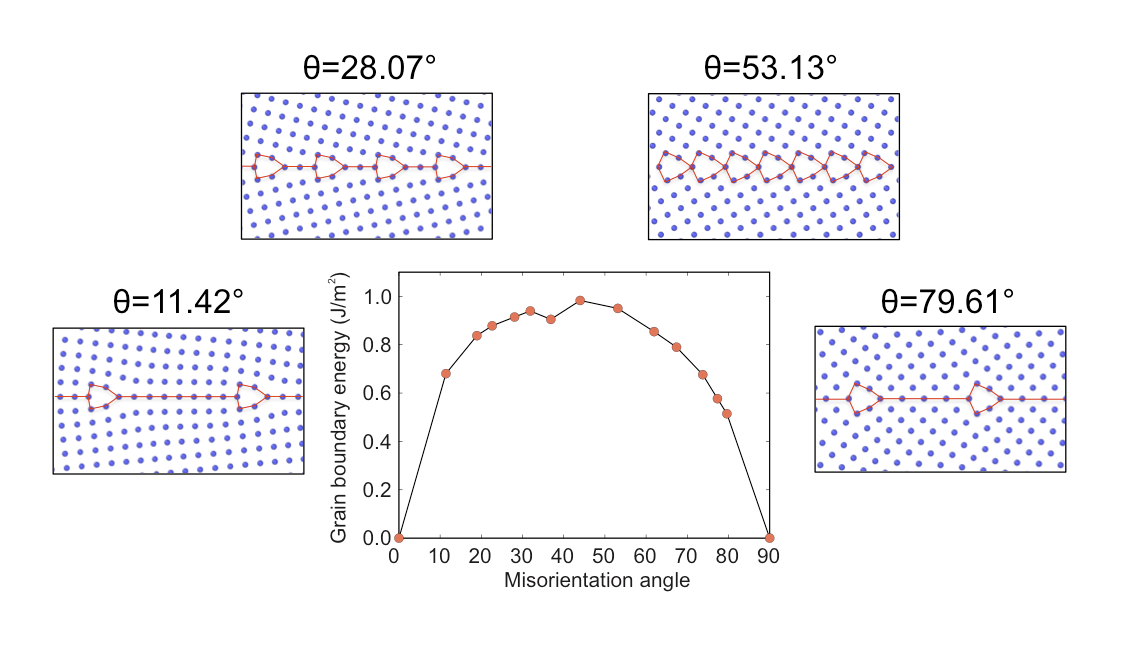} 
\par\end{centering}

\protect\protect\caption{\textbf{Grain boundary structures predicted by the conventional methodology
.} The kite family of grain boundary structures predicted by the conventional
simulation methodology. Low-energy grain boundary structures of symmetric
tilt boundaries in fcc Cu are composed of Kite shaped structural units.
The units separation distance changes with the misorientation angle
$\theta$. The construction does not add or remove atoms from the
grain boundary core, so not all possible states are sampled. Grain
boundary energy of the 13 grain boundaries studied is plotted as a
function of misorientation angle. \label{fig:Kites_gamma_surface_approach}}
\end{figure}

Figure~\ref{fig:Kites_gamma_surface_approach} illustrates the well-known
lowest energy configurations obtained by this approach\cite{Sutton83,Balluffi95}.
The structures of the boundaries are composed of kite shaped units.
The distance between these structural units depends on the misorientations
angle $\theta$. In the paper we will refer to this family of grain
boundary structures as the Kites family. For low-angle boundaries
composed of a periodic array of well-separated edge dislocations,
the kite-shaped units represent the dislocation core structure. Figure~\ref{fig:Kites_gamma_surface_approach}
illustrates grain boundary energy as a function of misorientation
angle $\theta$ obtained from the $\gamma$-surface construction.
This conventional methodology generates a large number of distinct
grain boundary states with different energies that correspond to different
grain translation vectors. However, all are built out of the same
fixed number of atoms compatible with the number of atoms in one plane
in each of the adjacent crystals. Due to this constraint many potentially
lower energy structures that have different atomic density are not
sampled\cite{Phillpot1992,Phillpot1994,Alfthan06,Chua:2010uq}.

On the other hand, the evolutionary search implemented in this work
samples very different grain boundary configurations by rearranging
atoms within a grain boundary core prior to relaxation, adding and
removing atoms from the boundary and changing the dimensions of the
grain boundary area on the fly. In a typical search several thousand
configurations are generated and their energy is evaluated using empirical
force fields. The low-energy configurations are automatically stored
and used later for the post-analysis.

\begin{figure}
\begin{centering}
\includegraphics[width=0.8\paperwidth]{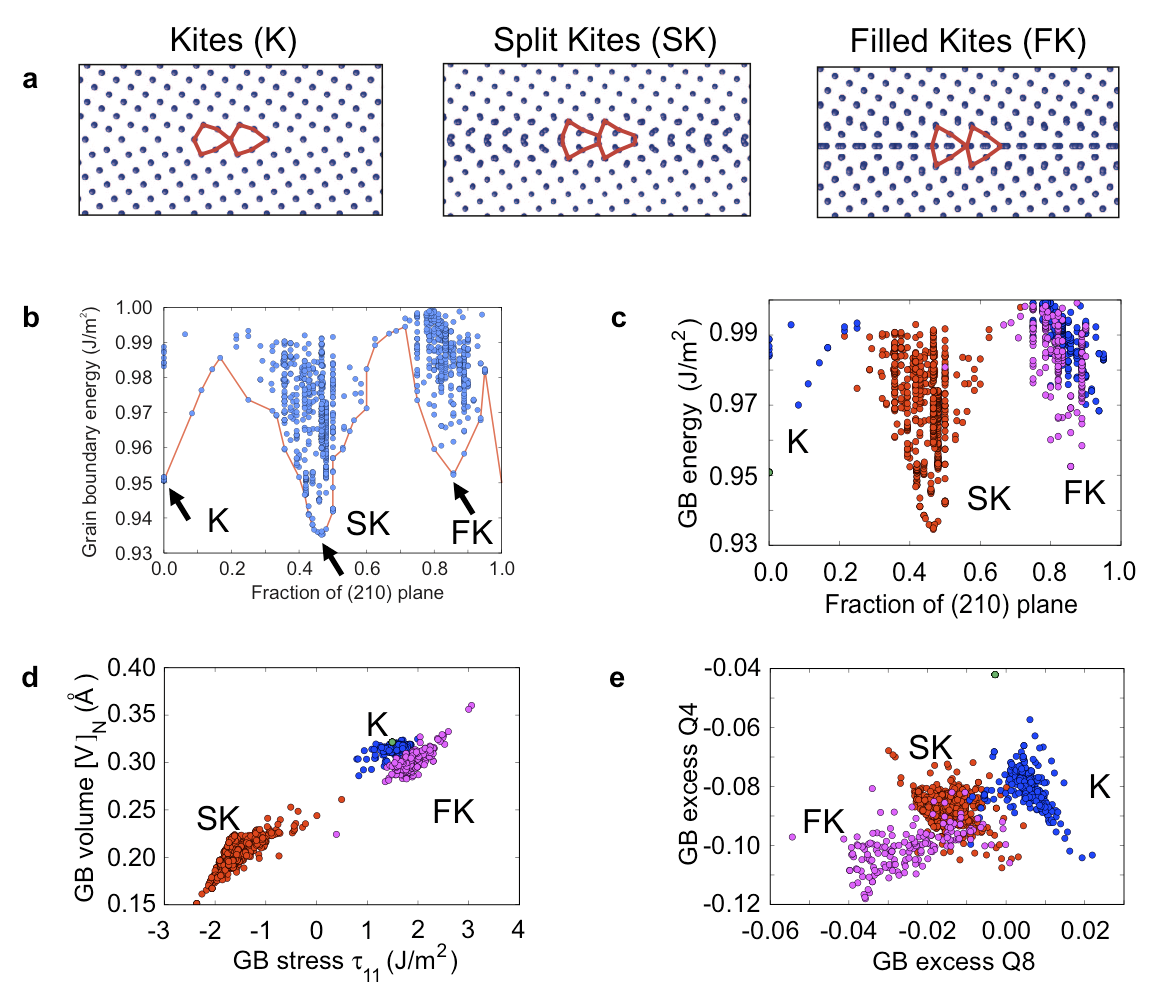} 
\par\end{centering}

\protect\protect\caption{\textbf{Evolutionary search and clustering analysis identify grain
boundary phases.} The evolutionary search and clustering analysis
identify three grain boundary phases of $\Sigma5(210)[001]$. The
evolutionary algorithm explores different atomic densities and identifies
multiple grain boundary phases: a) Kites, Split Kites and Filled Kites.
The three phases correspond to the energy minima as a function of
number of atoms. b) Energy of grain boundary configurations generated
by the evolutionary search as a function of number of atoms. d) and
e) The generated structures are automatically clustered into three
grain boundary phases according to similarities in their excess properties.
c) Grain boundary energy plot same as in b) with data points colored
according to the clustering. \label{fig:210Clustering}}
\end{figure}

A typical result of the evolutionary search for a $\Sigma5(210)[001]$
grain boundary is illustrated in Fig.~\ref{fig:210Clustering}b.
Because atoms are added and removed from the grain boundary core during
the search, the grain boundary energy of different configurations
is plotted as a function of the number of atoms in the system, which
is measured as a fraction of the number of atoms in a (210) plane.
Each point on the plot represents one particular structure generated
by the algorithm. The red line connecting the lowest energy configurations
for different atomic fractions shows that the grain boundary energy
has three distinct minima corresponding to different GB phases called
Kites, Split Kites and Filled Kites shown in Fig.~\ref{fig:210Clustering}a.
Prior modeling work demonstrated fully reversible transitions between
these different grain boundary phases \cite{Frolov2013,PhysRevB.92.020103,Frolov2016}.
The well-known Kite phase of this grain boundary is composed of the
structural units discussed earlier. The structures of the other two
phases on the other hand are more complex and are composed of multiple
distinct structural units. This structural diversity apparently gives
rise to a rich variety of low-energy Split Kite and Filled Kite configurations
that have different grain boundary dimensions. For example, nearly
degenerate in energy, but distinct Split Kite type structures were
found for cross-section sizes ranging from 1 to 25 times of the area
of the regular Kites (See Supplementary Fig.~4). This configurational
diversity should contribute to entropy of these grain boundary phases\cite{Han:2016aa,Han2017,doi:10.1080/01418610208240018}
and may have consequences for their high-temperature stability.

\bigskip{}

\noindent \textsf{\textbf{Clustering analysis.}} The three energy
minima shown in Fig.~\ref{fig:210Clustering}a represent the lowest
energy configurations of the three grain boundary phases. Other structures
generated by the evolutionary search may correspond to variations
of these three phases or belong to other grain boundary phases that
have not been identified yet. For example, a Kite configuration with
a single vacancy or an interstitial will have a different atomic density
and energy from that of the perfect Kite structure. However, this
defective grain boundary should still be identified with the Kite
phase. In general, each grain boundary structure generated by the
evolutionary search represents just one \textit{microstate}. A grain
boundary phase on the other hand is a \textit{macrostate}: it is represented
by an ensemble of similar micro-states. To identify distinct macro-states,
i.e. predict the number of grain boundary phases, we cluster the generated
grain boundary structures based on the similarity in their properties.
In a single component system a grain boundary is described by a set
of excess properties such as excess volume per unit area $[V]_{N}$,
grain boundary stress tensor $\hat{\tau}^{N}$ and number of atoms
$[n]$ (See Supplementary Note 2 for the definitions). First-order
phase transitions manifest themselves by discontinuous changes in
thermodynamic properties, which in turn suggests that these properties
could be used to distinguish different macrostates. In addition to
these thermodynamic properties which explicitly enter the equation
of state or the adsorption equation\cite{Gibbs,Cahn79,Frolov2012a},
we can formally introduce other excesses based on structural order
parameters. In this work we use Steinhardt order parameters $Q_{4}$,
$Q_{6}$, $Q_{8}$ and $Q_{12}$ designed to distinguish different
bulk phases based on local environments\cite{PhysRevB.28.784,doi:10.1063/1.2977970}.
In our work Q-series were calculated for each atom in the system and
the excess grain boundary amounts of $[Q_{i}]_{N}$ per unit area
were computed as described in the Supplementary Note 2. This new application
of the Q-series was developed to capture differences in local environment
present in different GB phases. We assign a vector $f=([n],[V]_{N},\tau_{N}^{11},\tau_{N}^{22},[Q_{4}]_{N},[Q_{6}]_{N},[Q_{8}]_{N},[Q_{12}]_{N})$
composed of four thermodynamic and four structural features to each
grain boundary configuration. A distance between two grain boundary
structures $a$ and $b$ is then calculated as
\[
d(f^{a},f^{b})=\sum_{i=1}^{8}((f_{i}^{a}-f_{i}^{b})/(f_{i}^{a}-f_{i}^{b}){}_{\text{max}})^{2}
\]
where all the feature differences were renormalized, so that their
values are in the range from 0 to 1. With the distance defined, the
clustering was performed using the method of fast search and find
of density peaks\cite{Rodriguez1492}. In this method for each data
point we calculate the number of neighbors $\rho_{i}$ within a cutoff
distance $d_{c}$ and the minimum distance $\delta_{i}$ from the
point to the other point that has a higher number of neighbors. The
centers of the clusters are then identified as points that have high
number of neighbors and separated from each other by the largest distances.
All other data points are then assigned to the closest cluster centers
which completes the clustering procedure. 

\bigskip{}

\noindent \textsf{\textbf{Clustering results for the $\Sigma5(210)[001]$
grain boundary.}} Figure~\ref{fig:210Clustering} illustrates an
example of the clustering analysis performed for the $\Sigma5(210)[001]$
boundary, which predicts three different grain boundary phases. To
visualize the data in the eight-dimensional space of the features
we show the data points projected on a plane formed by two different
excess properties. Figure~\ref{fig:210Clustering}c reveals strong
clustering of the data points based on properties such as excess volume
$[V]_{N}$ and excess stress $\tau^{N}$. The structures in the red
cluster were identified with Split-Kite phase, while the blue and
magenta represented Kites and Filled Kites, respectively. Note that
the Split-Kite structures have properties very different from both
Kites and Filled-Kites. On the other hand, Kites and Filled-Kites
phases have relatively similar thermodynamic properties and the excess
properties based on order parameters proved useful to distinguish
the two phases as shown in Fig.~\ref{fig:210Clustering}d. Overall,
Fig.~\ref{fig:210Clustering} demonstrates that clustering based
on multiple GB excess properties can be used to identify distinct
grain boundary phases. The analysis also reveals the degree to which
the thermodynamic properties can vary within each macro-state, which
provide insights regarding the stability of the different grain boundary
phases.

\bigskip{}

\begin{figure}
\begin{centering}
\includegraphics[height=0.6\paperheight]{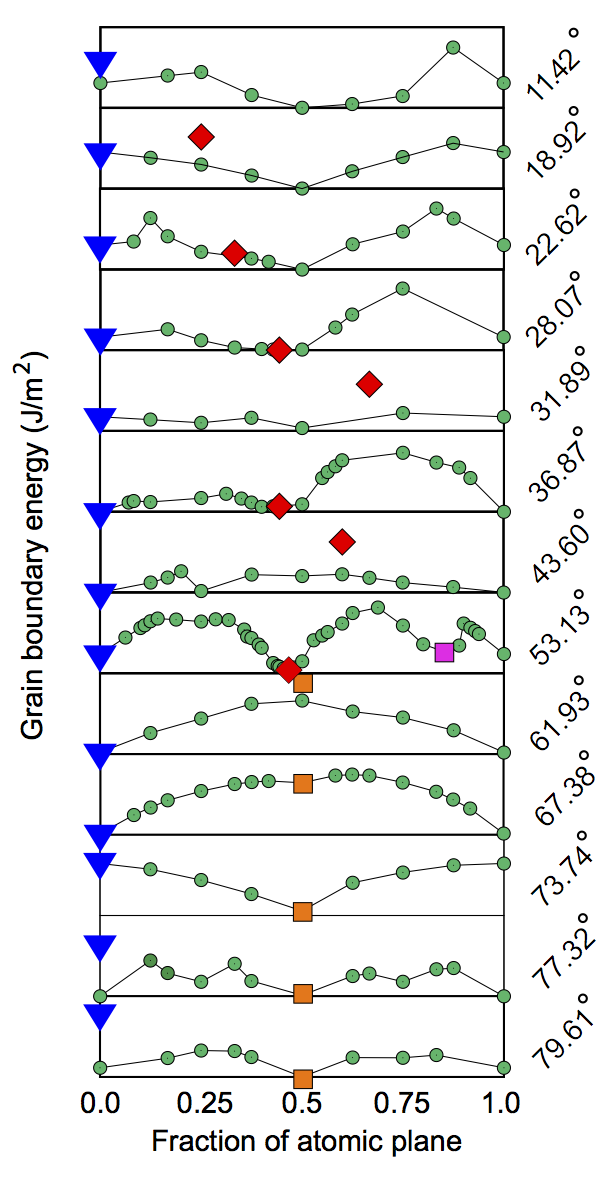} 
\par\end{centering}

\protect\protect\caption{\textbf{Energy map of grain boundary phases.} Evolutionary search
and clustering identify new ground states and multiple grain boundary
phases. The search explores different atomic densities and finds low-energy
grain boundary configurations (green circles) ignored by the conventional
methodology. For each grain boundary (angle $\theta$) atomic fractions
and energies of different grain boundary phases at 0 K are indicated
by blue triangles (Kite family) , red diamonds (Split Kite family)
and orange squares (Extended Kite family). \label{fig:uspex energy}}
\end{figure}

\noindent \textsf{\textbf{Grain boundary energy as a function of angle
$\theta$ and atomic density.}} In contrast to the $\gamma$-surface
construction which assumes that grain boundary energy is a function
of misorientation angle $\theta$ alone, the evolutionary search and
the clustering analysis of the $\Sigma5(210)[001]$ boundary demonstrates
the importance of exploring different atomic densities. In this work
we reconstruct GB energy as a function of the misorientation angle
and number of atoms in the boundary core. Figure~\ref{fig:uspex energy}
illustrates the results of the grand-canonical search spanning the
entire misorientation range of symmetric tilt boundaries from $0^{\circ}$
to $90^{\circ}$. For each of the 13 grain boundaries studied, the
green curves on the plot show the lowest GB energy calculated versus\ the
atomic fraction of the corresponding grain boundary plane. The blue
triangles at the origin of the plot correspond to the Kite structures
obtained by the $\gamma$-surface approach that does not add or remove
atoms. The plot reveals that within the entire misorientation range
the evolutionary search finds new ground states that require a change
in the atomic density. Most boundaries within two angle intervals
$0^{\circ}<\theta<53.13$ and $73.74^{\circ}<\theta<90.0^{\circ}$
exhibit at least one strong minimum which is close to about half of
the atomic plane fraction. These two intervals are separated by a
narrow range of angles around $65^{\circ}$ where the grain boundary
structures with unconventional density become unfavorable at 0~K.
This interval separates grain boundary groups with different structural
units. Many boundaries especially in the high-angle range exhibit
multiple minima suggestive of multiple grain boundary phases. There
are yet other boundaries with misorientation angles of $31.89^{\circ}$
and $43.60^{\circ}$ that show almost negligible variation in energy
with changing atomic density. This behavior suggests that these boundaries
can absorb point defects with no energetic penalty and may not be
very stable against fluctuation of atomic density.

Low-angle boundaries near the $0^{\circ}$ and $90^{\circ}$ are composed
of periodic arrays of isolated edge dislocations. The evolutionary
search results shown in Fig.~\ref{fig:uspex energy} indicate that
the dislocation core structure can be represented by multiple atomic
configurations that generally also require grand-canonical optimization:
atoms have to be added or removed from the dislocation core. The multiple
dislocation core configurations are examples of 1D phases, referred
in recent literature as 1D complexions\cite{Kuzmina1080,Wang:2014aa}.
Different core structures and transitions may have a strong effect
on dislocation mobility\cite{Kaplan1059}.

\begin{figure}
\begin{centering}
\includegraphics[height=0.6\paperheight]{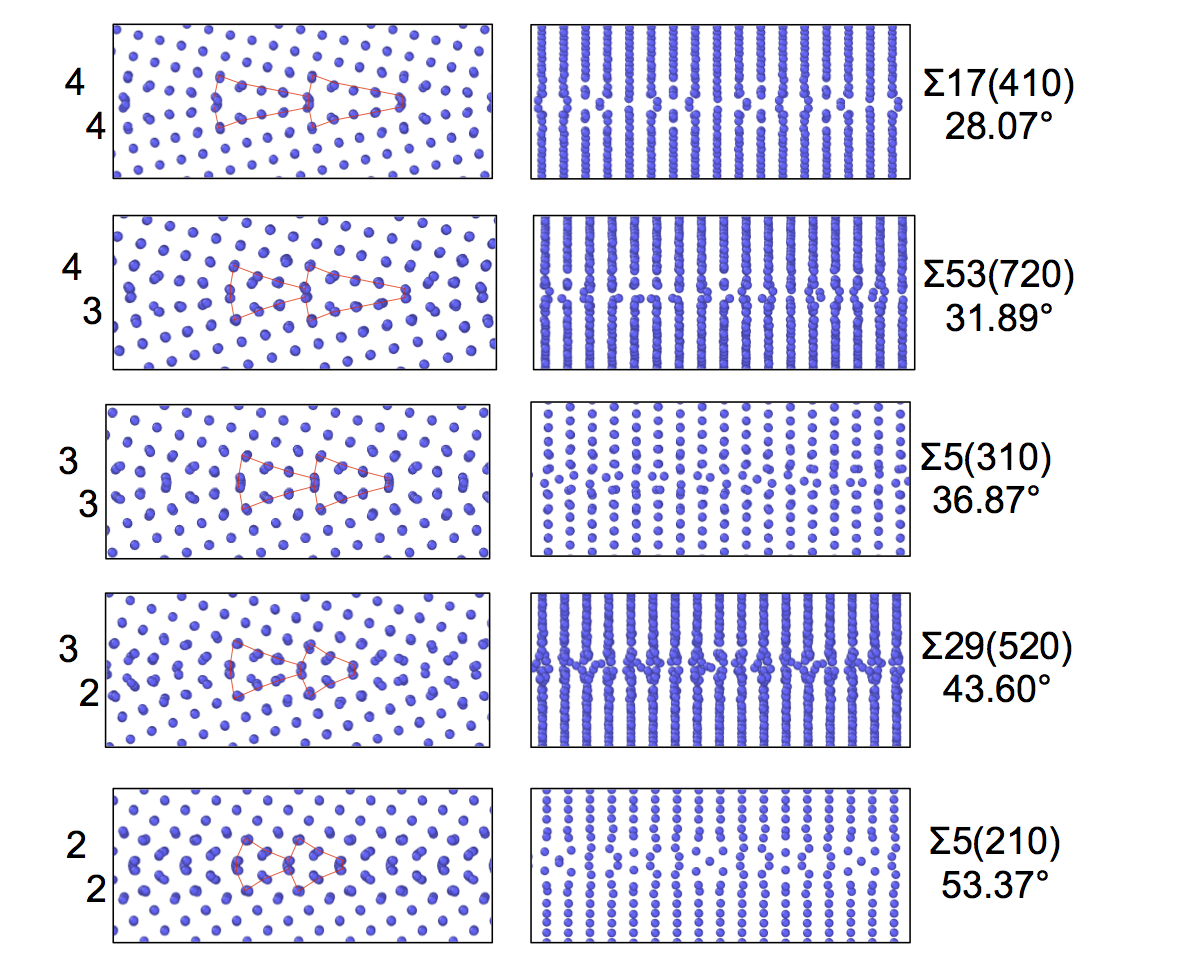} 
\par\end{centering}

\protect\protect\caption{\textbf{Split Kite family.} Split Kite phases of five representative
boundaries predicted by the evolutionary search and clustering analysis.
For each misorientation GB structures are viewed parallel to the {[}001{]}
tilt axis (left column) and normal to it (right column). The size
of the structural units of this family changes with misorientation.
Split Kites have higher atomic density compare to Kites, with extra
atoms occupying interstitial positions between {[}001{]} planes. \label{fig:uspex-SK-family}}
\end{figure}

Despite the large number of new grain boundary configurations found,
this richness of structures is easy to comprehend because they can
be grouped into families of structures with similar characteristic
units. The Kite family illustrated in Fig.~\ref{fig:Kites_gamma_surface_approach}
was already introduced with the $\gamma$-surface approach and has
different grain boundaries with similar kite-shaped structural units
and atomic density. Our grand canonical evolutionary search identifies
two new families of grain boundary phases which we call Split Kites
and Extended Kites. In the energy vs. atomic density map in Fig.~\ref{fig:uspex energy}
the three families are indicated by blue triangles (Kites), red diamonds
(Split Kites) and orange squares (Extended Kites). Fig.~\ref{fig:uspex-SK-family}
illustrates split kite structures for several representative boundaries,
which are composed of similar structural units. Differently from Kites,
instead of changing the unit separation distance with changing the
misorientation angle, it is the size of the structural units that
changes with $\theta$. Fig.~\ref{fig:uspex-SK-family} illustrates
how the grain boundary structure of Split Kites changes when the misorientation
angle increases from $\theta=28.07^{\circ}$ to $\theta=53.37^{\circ}$.
$\Sigma17(410)[001]$ at $\theta=28.07^{\circ}$ is composed of units
with size equal to four $1/2[100]$ lattice spacings. $\Sigma53(720)[001]$
which has a higher misorientation of angle of $31.89^{\circ}$ consists
of alternating units with sizes 3 and 4. Other grain boundaries are
composed of units with size 3 only, alternating 3 and 2, until at
$\theta=53.37^{\circ}$ the $\Sigma5(210)[001]$ boundary is composed
of units with size 2. The $\Sigma5(210)[001]$ also exists in a Filled
Kite structure, which was not found in other 13 boundaries and is
likely to be stable in a narrow misorientation angle range around
$53.37^{\circ}$.

All Split Kite structures are characterized by higher atomic density
relative to Kite family. In the Kite family all the atoms at the boundary
are confined to the {[}100{]} planes. On the other hand, in all Split-Kite
structures additional atoms densely occupy positions in-between the
{[}001{]} planes, creating complex structures composed of multiple
distinct subunits. The atomic arrangement with the boundaries along
the tilt axis is illustrated in right-hand side of Fig.~\ref{fig:uspex-SK-family}.
This internal structure gives rise to a rich configurational diversity
and may contribute to the entropy of these structures at finite temperature.
Notice that in Fig.~\ref{fig:uspex energy} Split-Kite configurations
were not identified with the ground states for some misorientations,
see Supplementary Fig.~2 for further discussion on symmetries of
these structures. 

\begin{figure}
\begin{centering}
\includegraphics[width=0.6\paperwidth]{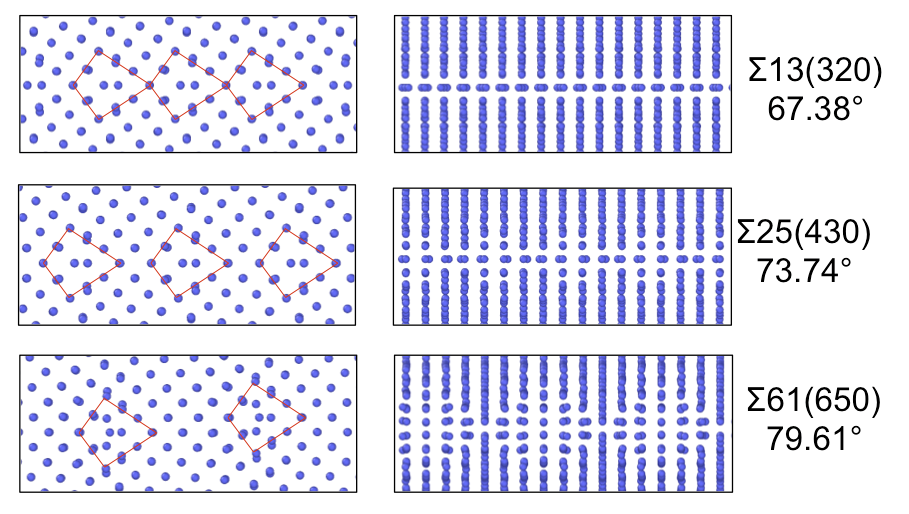} 
\par\end{centering}

\protect\protect\caption{\textbf{Extended Kite family}. Extended Kite phases of three representative
boundaries predicted by the evolutionary search at 0 K. The misorientation
angles are indicated on the figure. For each misorientation GB structures
as viewed parallel to the {[}001{]} tilt axis (left column) and normal
to it (right column). Extended Kites have higher atomic density compare
to Kites, which correspond to half of the atomic plane. The structural
units are outlined and change their separation with the increasing
misorientation angle. \label{fig:uspex-EX-family-1}}
\end{figure}

Different structural units appear at misorientation angles $\theta>53.37^{\circ}$
and are illustrated in Fig.~\ref{fig:uspex-EX-family-1}. The units
of the boundary are {[}110{]} edge dislocations with more extended
dislocation core structure than regular Kites. For this reason we
refer to this family of grain boundaries as Extended Kites. Similar
to Split Kites, the Extended kites are denser than Kites and become
more energetically favorable as the misorientation angle $\theta$
increases away from $61.93{}^{\circ}$. The misorientation interval
$53.37^{\circ}<\theta<61.93{}^{\circ}$ represents a transition region
where both structural units may be equally favorable at some temperature.
Grain boundary structure in this misorientation range is likely to
exhibit checkerboard pattern composed of both split-kites and extended
kites structural units\bigskip{}

\begin{figure}
\begin{centering}
\includegraphics[width=0.7\paperwidth]{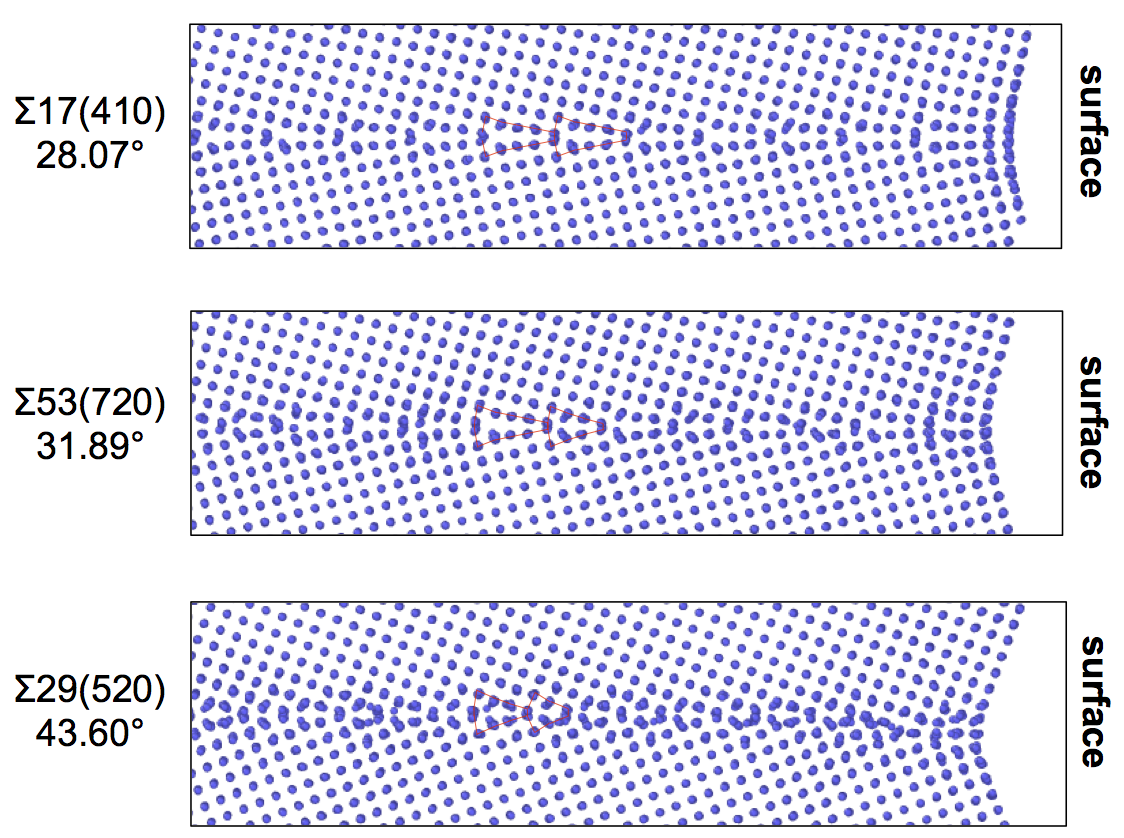} 
\par\end{centering}

\protect\protect\caption{\textbf{Equilibrium structures at high temperature. }High-temperature
Split-Kite grain boundary phases of three representative grain boundaries
independently predicted by molecular dynamic simulations. Grain boundary
phase transitions occur in the simulations, because open surfaces
and grain boundary diffusion at 900 K enable variation of the atomic
density in the GB core. The structures match the predictions of the
evolutionary search at 0 K. \label{fig:highT-MD}}
\end{figure}

\noindent \textsf{\textbf{Grain boundary structures and transitions
at finite temperature.}} To validate the structures predicted at 0~K
by the evolutionary search and demonstrate possible grain boundary
phase transitions, we performed high-temperature MD simulations of
a subset of relatively high-angle boundaries. In these simulations
the grain boundaries were terminated at open surfaces following the
methodology proposed in Ref.~~\cite{Frolov2013}. Open surfaces
act as sources and sinks of atoms and effectively introduce grand-canonical
environment in the grain boundary core. This approach is less effective
for low-angle boundaries due to much lower diffusivity normal to the
tilt axis. We chose regular kite structures illustrated in Fig.~\ref{fig:Kites_gamma_surface_approach}
as the initial configurations prior to annealing. During the 900~K
anneal for tens of nanoseconds the grain boundaries transformed to
Split Kite configurations. Figure~\ref{fig:highT-MD} illustrates
three representative high-angle grain boundaries following the transformation.
The high-temperature structure of these boundaries matches Split Kite
configurations independently generated by the evolutionary search.
These MD simulations show that the Split Kite family represents the
structure of grain boundaries at high temperature and confirm that
our structure sampling at 0~K can generate grain boundary phases
relevant to finite temperature.

\begin{figure}
\begin{centering}
\includegraphics[width=0.75\paperwidth]{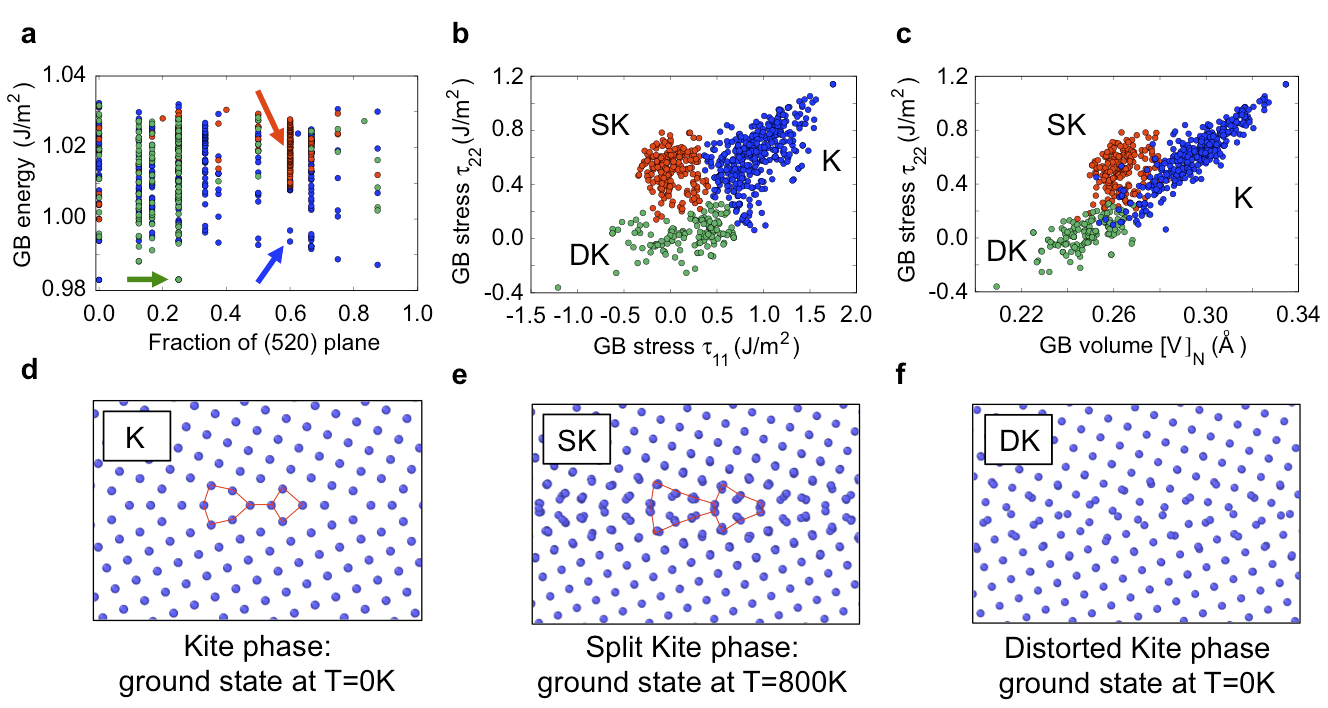} 
\par\end{centering}

\protect\protect\caption{\textbf{Clustering identifies grain boundary phases observed at high
temperature. }Metastable grain boundary phases identified by the clustering
analysis at 0 K become stable at high temperature. Evolutionary search
a) and clustering b), c) for the $\Sigma29(520)[001]$ predict three
grain boundary phases: d) Kites, e) Split Kites and f) Distorted Kites.
Kite phase is the ground state at 0~K. Split Kite phase is a high
energy state at 0 K, but becomes a ground state with the lowest free
energy at finite temperature, as demonstrated by a transformation
in MD simulations at 900 K. \label{fig:clustering520}}
\end{figure}

For a number of misorientations we find that that Split Kite structures
observed in high temperature MD simulations are not the ground state
at 0~K as illustrated in Fig.~\ref{fig:uspex energy}. These grain
boundaries have different structures at low and high temperature and
exhibit first-order transitions that result in discontinuous changes
in properties, analogous to those reported in the recent experimental
studies\cite{Divinski2012}. For example, $\Sigma5(210)[001]$ and
$\Sigma5(310)[001]$ exhibit such transitions\cite{Frolov2013} and
the different GB phases are easy to identify even at 0~K because
they correspond to distinct GB energy minima as a function of number
of atoms. On the other hand, in some boundaries such as $\Sigma29(520)[001]$
and $\Sigma53(720)[001]$, Split Kite structures do not correspond
to such minima and cannot be found within the lowest energy configurations
at 0~K. In this case, the clustering analysis becomes invaluable
for the identification of the potential high-temperature grain boundary
phases.\bigskip{}
\textsf{\textbf{Clustering results for the $\Sigma29(520)[001]$ grain
boundary.}} Fig.~\ref{fig:clustering520}a illustrates results of
the energy search generated for the $\Sigma29(520)[001]$ GB by the
evolutionary algorithm. Notice that the energy as a function of atomic
density shows no obvious minima, like the minima observed for $\Sigma5(210)[001]$,
so it is not clear from this plot alone that this boundary may have
multiple phases. Figure panels \ref{fig:clustering520}b and c shows
the excess properties of the generated structures and the clustering
analysis identifies three distinct phases. The representative grain
boundary structures from the three different clusters are illustrated
in Figs.~\ref{fig:clustering520}d-f. The red cluster of points corresponds
to Split Kite configuration shown in Fig.~\ref{fig:clustering520}e
and observed at high temperature. The majority of the configurations
has atomic fraction of 0.6. The energy plot in Fig.~\ref{fig:clustering520}a
clearly demonstrates that Split Kites represent higher energy state
compare to all other configurations even within the subset with the
atomic fraction of 0.6. This clustering demonstrates that the examination
of the lowest energy configurations alone is not sufficient and will
fail to predict the high-temperature GB phases. The clustering analysis
captures the heterogeneity in properties of the generated structures
and identifies multiple macrostates. Some macro-states may not be
the lowest energy configurations at 0~K, but can be potentially become
the lowest free energy state at finite temperature or with varying
chemical composition. The evolutionary search and clustering analysis
complemented by energy calculations can generate grain boundary phase
diagrams and predict grain boundary phase transitions.

\bigskip{}

\noindent \textsf{\textbf{\Large{}Discussion}}{\Large \par}

Using the advanced evolutionary sampling and clustering analysis we
have uncovered rich phenomena unexplored by previous computational
studies of grain boundaries. Based on the successes of applying evolutionary
algorithm in the prediction of bulk crystals, surfaces and clusters,
we developed a computational tool to explore the low-energy GB structures
in a vast compositional, dimensional, and structural space. To address
the challenges of explosive increased searching space in large systems,
we followed the idea of coarse-grained modeling and define the simplified
structure representations in the evolutionary algorithm. The developed
algorithm generates a diverse population of configurations while adding
and removing atoms from the grain boundary core and changing grain
boundary dimensions. In this work the evolutionary search was applied
to reconstruct grain boundary energy surface as a function of both
misorientation and atomic density in a model system of Cu symmetric
tilt boundaries and predicted new ground states of grain boundaries
within the entire misorientation range. For most misorientations multiple
grain boundary phases were found demonstrating that phase behavior
of interfaces is a general and common phenomena, not limited to few
special high-angle boundaries.

The computational discovery of these phases and modeling of the transitions
became possible only with the new methodology. Specifically, we designed
a clustering procedure that analyses the results of the evolutionary
search and automatically identifies different macro-states or grain
boundary phases by grouping the individual configurations according
to their thermodynamic and symmetry properties. While many studies
of structure prediction at 0~K often focus on finding configurations
with the lowest energy possible, the clustering analysis examines
grain boundary structures within a finite energy interval and identifies
multiple metastable grain boundary phases in addition to the ground
state. While for some misorientations these metastable states were
also the energy minima as a function of atomic density, in general
they are just higher energy macro-states that are not minima of energy
as a function any particular property and as such were identified
only with help of the clustering analysis.

High-temperature MD simulations with open surfaces demonstrated first-order
grain boundary transitions between the different grain boundary phases
independently predicted by 0~K calculations. This confirms that the
ground states and metastable states generated by the evolutionary
search and the clustering analysis at 0~K are relevant to prediction
of grain boundary structures at finite temperature. Moreover, in principle
the temperature induced grain boundary phase transitions can be predicted
by calculating the free energy of the different metastable states
using available computational methods\cite{PhysRevLett.114.195901,PhysRevLett.100.020603,doi:10.1063/1.4945653,FREITAS2016333,PhysRevB.95.155444}.
Thus in the future, the 0~K search developed in this work augmented
with an efficient free energy calculation scheme can be used to construct
grain boundary phase diagrams.

In this work we demonstrate that within the entire misorientation
range certain types of structures with similar characteristics can
be grouped into families of Kites, Split-Kites and Extended-Kites.
For example, the characteristic features of the Split-Kite phase is
their higher atomic density compared to that of Kites and configurationally
more diverse atomic arrangement of the structure. Split-Kites were
found to be the high-temperature phases for the majority of grain
boundaries studied. The presence of distinct families of grain boundaries
like Kites, Split-Kites and Extended-Kites with properties that are
different across the entire misorientation range may help explain
the sharp discontinuous transitions in mobility observed in polycrystalline
materials. For example, addition of impurities with large size mismatch
would stabilize Kite family of grain boundary structures over the
much denser Split-Kite and Extended-Kites families in the entire polycrystalline
sample. The ability to predict families of phases and their characteristic
excess properties may provide guidance on how interfaces with certain
structure and properties can be enforced in a material by alloying
elements or temperature, ultimately providing a way to achieve the
desired materials microstructure and properties.

The insights gained in this work about grain boundaries are also relevant
to other lattice defects such as dislocations and triple junctions.
Low-angle boundaries near 0 and 90 misorientations studied in this
work are composed of rows of edge dislocations. The evolutionary search
predicted new ground states of dislocation core structures. The optimization
required sampling of different atomic arrangements as well as addition
and removal of atoms form the dislocation core. This type of sampling
was not typically performed in studies that attempted to predict dislocation
structures. It is well known that the core structure can have a pronounced
effect on dislocation mobility. The systematic investigation of different
dislocation core structures and their properties is subject to future
work.

\bigskip{}

\noindent \textsf{\textbf{\large{}Methods}}{\large \par}

\noindent \textsf{\textbf{GB structure calculations at 0 K. }}For
each grain boundary we ran 3-5 independent evolutionary searches.
Each search evolves over up to fifty generations. The search explores
different atomic densities ranging from 0 to 1 measured as a fraction
of number of atoms found in one bulk atomic plane parallel to the
grain boundary. We conducted structure searches sampling the entire
range of densities as well as searches constrained around certain
atomic densities and found that both types of searches are useful.
A typical run explores the structures ranging from 500 to 5000 atoms
for the entire model and 30 to 300 atoms for the GB region. For each
grain boundary we explore different grain boundary areas by replicating
the smallest possible cross-section up to 25 times. See Supplementary
Note 1 for more details. The energy of the generated configurations
was evaluated with LAMMPS code\cite{Plimpton95}.

\bigskip{}

\noindent \textsf{\textbf{Finite temperature simulations.}} Molecular
Dynamics simulations were performed in the NVT ensemble with Nose-Hoover
thermostat using the LAMMPS code\cite{Plimpton95}. Periodic boundary
conditions were applied only along the {[}001{]} tilt axis. In the
direction normal the grain boundary plane the simulation block was
terminated by two boundary regions in which the atomic positions were
kept fixed during the simulation. In the $x$ direction the boundaries
were terminated by two open surfaces. The dimensions of the simulation
block were 50~$\textup{\AA}$ along the tilt axis and 200~$\textup{\AA}$
in the direction normal to the grain boundary plane. In the $x$ direction
the block size varied from 250 to 350 $\textup{\AA}$ depending on
the misorientation angle. Isothermal simulations at T=900~K (0.678
$T_{m}$) and T=800~K (0.602 $T_{m}$) were performed for 200 ns
each. 

\bigskip{}

\section*{\textsf{\large{}Acknowledgements}\textsf{ }}

This work was performed under the auspices of the U.S. Department
of Energy by Lawrence Livermore National Laboratory under Contract
No. DE-AC52-07NA27344. The work was funded by the Laboratory Directed
Research and Development Program at LLNL under project tracking code
17-LW-012. BL gratefully acknowledges support by the LLNL LDRD program
for this work. We acknowledge the use of LC computing resources. Work
at UNLV is supported by the National Nuclear Security Administration
under the Stewardship Science Academic Alliances program through DOE
Cooperative Agreement DE-NA0001982.

\section*{\textsf{\large{}Competing financial interests}}

The authors declare that they have no competing financial interests.

\newpage{}\clearpage{}

\bigskip{}

\bigskip{}

\bigskip{}

\newpage{}

\clearpage{}\setcounter{page}{1}

\title{\noindent \textsf{\textbf{\large{}Predicting phase behavior of grain
boundaries with evolutionary search and machine learning}}}

\title{\bigskip{}
}

\author{\noindent Qiang Zhu$^{1}$,\enskip{}Amit Samanta$^{2}$,\enskip{}Bingxi
Li$^{3}$, Robert E. Rudd$^{2}$ and Timofey Frolov$^{2}$}

\title{\bigskip{}
}

\lyxaddress{$^{1}$Department of Physics and Astronomy, High Pressure Science
and Engineering Center, University of Nevada, Las Vegas, NV 89154,
USA}

\lyxaddress{$^{2}$ Lawrence Livermore National Laboratory, Livermore, California
94550, USA}

\lyxaddress{$^{3}$ Department of Materials Science and Engineering, University
of California, Davis, CA 95616, USA}

\bigskip{}

\bigskip{}

\noindent \textbf{Supplementary Materials include:}

\noindent \textbf{Supplementary Notes 1 and 2, Figures S1, S2, S3
and S4}

\subsubsection*{Supplementary Note 1}

Evolutionary algorithms (EAs) adopts concepts from evolutionary biology
based on populations, selection, reproduction by heredity and mutation,
aimed to locate the individual with highest fitness. The code generates
a population of grain boundary structures and improves them over several
generations to predict low-energy configurations. During the evolution
complex and diverse structures with different atomic densities are
sampled by operations of heredity and mutation which involve atomic
rearrangements as well as addition and removal of atoms from the grain
boundary core. A pictorial representation of the evolutionary algorithm
scheme and a schematic illustration of the GB calculation are shown
in Fig. S1. In our implementation, we split each GB model into three
different regions, the region of upper grain (UG) and lower grain
(LG), and grain boundary (GB) . Our optimization target is the atomic
configuration in GB and the relative translation between UG and LG
leading to the lowest GB energy. UG and LG regions are pre-specified
(typically 40-60 $\textup{\AA}$ thick), which can be either tilt
or twist. Although only the symmetric tilt models are studied in this
work, this method could be applied to all types of GBs and interfaces.
The GB thickness is a pre-defined variable by the user. In order to
make sure the full convergence in GB energy calculation, we expand
GB region by adding the buffer zone from both UG and LG, approximately
20 $\textup{\AA}$ for each direction. We first randomly generate
the atomic coordinates in GB slabs with a random cross-section, and
random layer group symmetries, and then join UG-GB-LG together with
random translations between UG-GB, and GB-LG. The structures are then
relaxed by external computational codes either based on empirical
force fields or ab-initio calculations, followed by fitness evaluation,
namely, the excess GB energy in this case. The UG-GB-LG model should
be sufficient for simulation with periodic boundary conditions along
GB plane and open boundaries perpendicular to the GB plane. If only
3-dimensional periodic boundary condition are available in the ab-initio
codes, a vacuum layer of 10-20 $\textup{\AA}$ should be added on
top of UG, in order to eliminate the interaction between UG and LG.
During the geometry optimization, the atoms at GB (LG) region need
to be fully relaxed (fixed), while the atoms in the UG can only move
as whole by rigid body translation. Structures with better fitness
are more likely to be selected (according to tournament selection)
as parents to generate the new child structures in the following ways:
1) heredity which choses two GB structures and randomly slices them
at the same position in the GB unit cell and then combines the pieces
to generate the offspring; 2) mutation which choses one GB structure
and displaces its atoms according to the stochastically picked soft
vibrational modes based a bond-hardness model; 3) insertion/removal
of atoms, which choses one GB structure and randomly inserts or deletes
some atoms in the GB slab. The offspring, together with a few best
structures from the previous generation, comprise the new population.
This whole cycle is repeated until no lower-energy structures are
produced for sufficiently many generations. 

To remove atoms from the GB slab, the algorithm first calculates the
local order parameter for each atom in the region. The order parameter
is described in Eq. (5) of Ref. \cite{LYAKHOV20101623}. A random
fraction of atoms (not exceeding 25\%) with the lowest degree of order
is then deleted. To insert atoms into the GB slab, we identify sites
unoccupied by atoms by constracting a uniform grid with a resolution
of 1 \AA $^{3}$ and fill them at random. To ensure relatively gradual
changes in the GB structure, the random number of the inserted atoms
also does not exceed 25\% of the total number of atoms in the GB slab.
It should be noted that both insertion/removal and heredity operations
automatically involve the change of number of atoms at GB. 

GB structure might have rather complex and large-scale reconstructions.
Therefore, we allow the GB dimentions to vary automatically during
the search. For heredity and insertion/removal of atoms operations,
we first expand (or shrink) the parent structures to the new size.
For mutations, we calculate the atomic displacements corresponding
to both zero and nonzero wave vectors, enabling cell size to spontaneously
change during the simulation\cite{Zhu-PRB-2015}. By allowing the
number of GB atoms and the size of GB cells to vary in the course
of structural evolution, we can eliminate the unphysical constraints
in the traditional \textgreek{g}-surface approach, thus enabling a
more complete sampling. 

In this work, we enable the automated exploration in higher dimensional
space, which includes the structures with variable number of atoms
and variable cell sizes. A typical run would explore the structures
ranging from 500 to 5000 atoms for the entire model and 30 to 300
atoms for the GB region. It is well known that the complexity exponentially
increases with the growing dimensionality. In that case, a key to
ensure efficient sampling is to balance between individual quality
and population diversity. Any pure random structure initialization
or variation operation is very likely to lead to disordered, liquid-like
structures with close energetics. To address this challenge, we followed
the idea of coarse-grained modeling and define the simplified representations
during the stage of structure generation. Some key representations
used are symmetry, vibrational modes and degree of local order. In
order to predict very large systems, we made several key improvements
compared to other existing approaches. Although almost all genetic/evolutionary
algorithms are designed to start from random structures for the first
generation, a fully random initialization is a poor choice for large
systems. Symmetry has played a crucial role in the analysis of crystal
structures and recently been extended to grain boundaries. We proposed
a novel initialization scheme that only generates the structures with
the desired layer group (or space group). In the case that the truly
low-energy structures cannot be described by the symmetry, the symmetry
could be broken or lowered by the subsequent variation operations
like heredity and mutation. We apply mutation in a reduced variable
space: instead of displacing the atoms randomly or based on a Gaussian
distribution, we calculate the vibrational modes corresponding to
both zero and nonzero wave vectors and displace the atoms along those
soft modes (i.e., the vibrations with negative or small positive frequencies)\cite{Zhu-PRB-2015}.
The advantages are twofold. First, it mimics the structure transition
due to phonon instability upon large elastic strain, thus is more
likely to lead to child structure with low energy. Second, it naturally
enables the cell size to spontaneously change during the simulation
and thus could efficiently identify the optimum cell sizes due to
structural modulation. Degree of local order: Recently, the local
degree of order was introduced to characterize the quality of the
environment and its symmetry for a given atomic position in the structure\cite{Lyakhov-CPC-2013}.
This concept turns out very useful to evaluate the contribution of
each atom to the total energy. Therefore, it could serve as the basis
during the fragment selection, add and removal in the EA variation
operations. Atoms with higher order should have higher probability
to be selected and lower probability to be deleted. 

\begin{figure}
\begin{centering}
\includegraphics[width=1\textwidth]{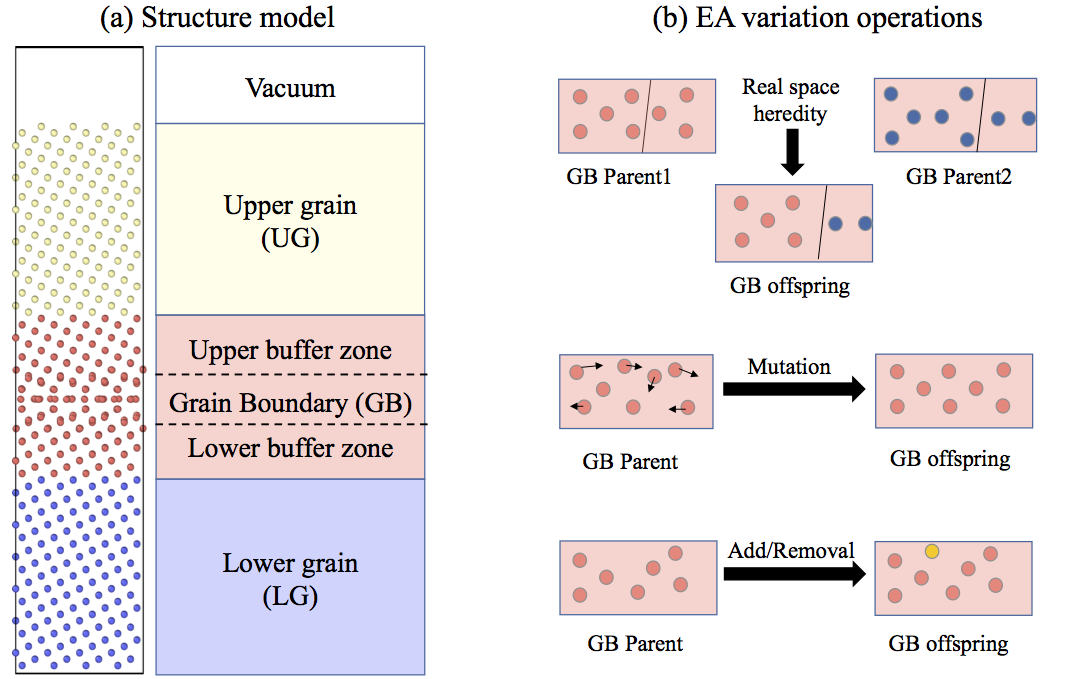} 
\par\end{centering}

\bigskip{}

\noindent \begin{centering}
\textbf{Supplementary Figure S1. }
\par\end{centering}

Schematic of the evolutionary algorithm (EA) for grain boundary prediction.
(a) the model of GB representation; (b) the scheme of variation operators
to generate new offspring in the context of EA. \label{fig:genetic_alg_flow_chart}
\end{figure}

\subsubsection*{Supplementary Note 2}

Each grain boundary structure was characterized by eight excess properties.
In a single component system grain boundary free energy $\gamma$
is given by\cite{Gibbs,Cahn79}

\[
\gamma A=E-TS-\sigma_{33}V-\mu N=[E]_{N}-T[S]_{N}-\sigma_{33}[V]_{N}
\]
where $[Z]_{X}$ are grain boundary excess properties expressed using
Cahn's determinants. Grain boundary free energy is a function temperature,
stress and lateral strain as described by the adsorptions equation\cite{Gibbs,Frolov2012a,Frolov2012b}

\[
d(\gamma A)=-[S]_{N}dT-[V]_{N}d\sigma_{33}+\tau_{ij}Ade_{ij,}\qquad i,j=1,2
\]
where $e_{ij}$ is the elastic strain tensor. At 0~K we calculate
excess volume $[V]_{N}$ and two components of grain boundary stress
$\tau_{11}$ and $\tau_{11}$ as

\[
[V]_{N}=\frac{1}{A}(V-V^{\text{{bulk}}}N/N^{\text{{bulk}}})
\]

\[
\tau_{11}=[\sigma_{11}V]_{N}=\frac{1}{A}(\sigma_{11}V-\sigma_{11}^{\text{{bulk}}}V^{\text{{bulk}}}N/N^{\text{{bulk}}})
\]

\[
\tau_{22}=[\sigma_{22}V]_{N}=\frac{1}{A}(\sigma_{22}V-\sigma_{22}^{\text{{bulk}}}V^{\text{{bulk}}}N/N^{\text{{bulk}}})
\]
Notice that $V^{\text{{bulk}}}/N^{\text{{bulk}}}=\Omega$ is a volume
per atom in the bulk. In atomistic simulations volume occupied by
each atom was calculated by LAMMPS using the Voronoi construction\cite{Plimpton95}.
The product $\sigma_{ij}V$ for each atom is also calculated by LAMMPS.
In our calculations bulk stresses are zero within the numerical accuracy. 

Another feature that we use to compare different grain boundary structures
is the quantity $[n]_{N}$ which we refer to as grain boundary atomic
density\cite{Frolov2013}. This quantity is fundamentally different
from excess volume $[V]_{N}$ and the two should not be confused.
First, we calculate the total number of atoms in the system $N$ and
the number of atoms $N_{\text{{plane}}}^{\text{{bulk}}}$ in one atomic
plane parallel to the GB and located inside the bulk part in the same
system. $[n]_{N}$ is then calculated as the ratio $Modulo(N_{\text{{plane}}}^{\text{{bulk}}},N)/N_{\text{plane}}^{\text{{bulk}}}$.
Since it is measured as a fraction of $N_{\text{{plane}}}^{\text{{bulk}}}$,
its value goes from 0 to 1. $[n]_{N}$ is also a periodic quantity:
addition of a complete plane results in return to the same grain boundary
structure. As a result, the atomic density distance between two structures
$a$ and $b$ was calculated as $min(abs([n^{a}]_{N}-[n^{b}]_{N}),1-abs([n^{a}]_{N}-[n^{b}]_{N}))$. 

In addition to the four features described above we introduced grain
boundary excess amounts of Steinhardt order parameters Q4, Q6, Q8
and Q12\cite{PhysRevB.28.784}. These parameters per atom are calculated
within LAMMPS\cite{Plimpton95}. The grain boundary excess amounts
of these parameters per unit area are then introduced in a manner
analogous to the thermodynamic excess properties. 

\[
[Q]_{N}=\frac{1}{A}(Q-Q^{\text{\text{bulk}}}N/N^{\text{bulk}})
\]
where $Q={\displaystyle {\displaystyle \sum_{i=1}^{N}Q^{i}}}$ is
the total amount of the order parameter in a region enclosing the
grain boundary and containing $N$ atoms, $Q^{\text{bulk}}/N^{\text{bulk}}$
is the value of this order parameter per atom in the bulk. Q is one
of the Q4, Q6, Q8 or Q12. 

\begin{figure}
\begin{centering}
\includegraphics[width=1.2\textwidth]{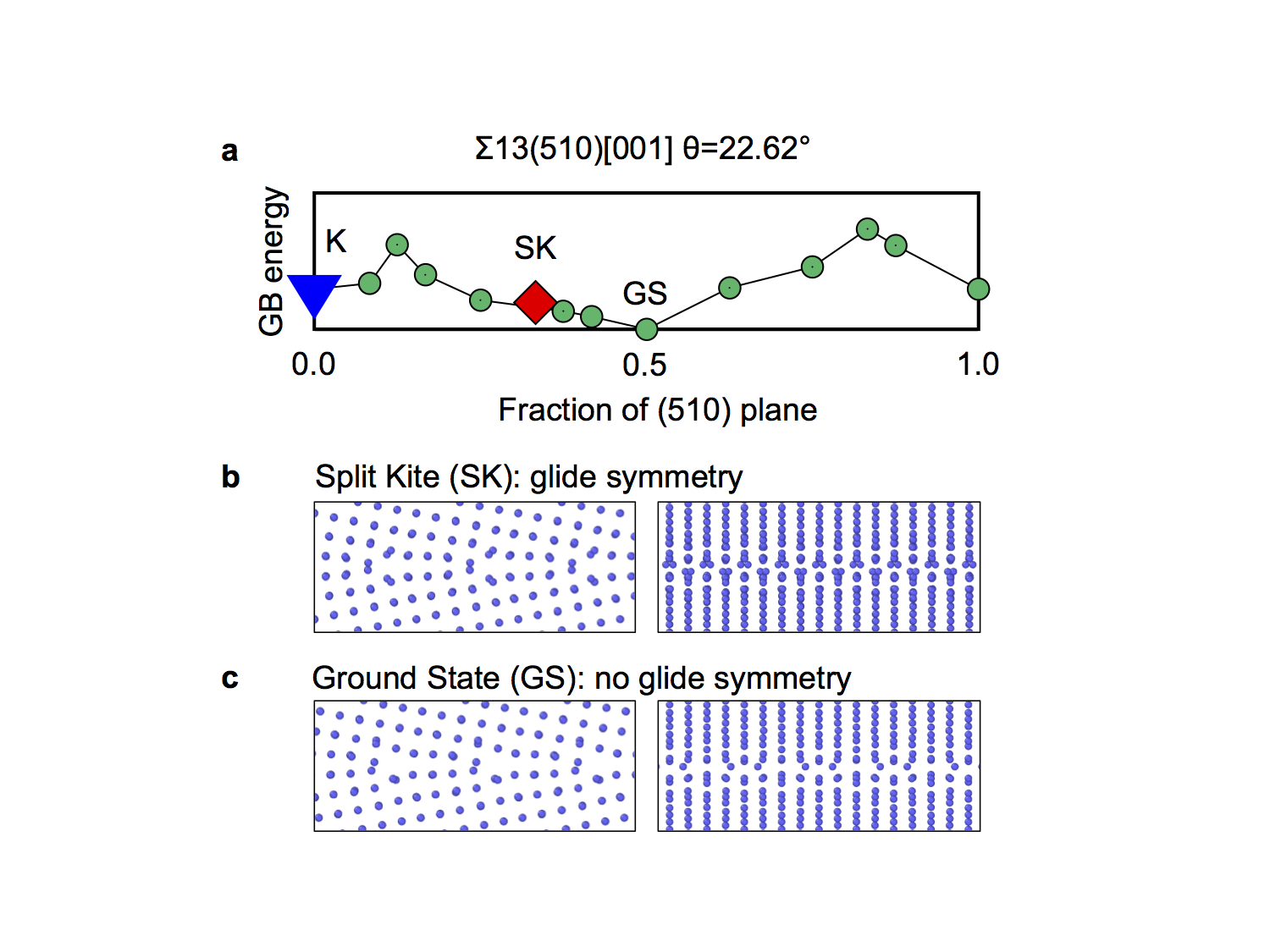} 
\par\end{centering}

\bigskip{}

\noindent \begin{centering}
\textbf{Supplementary Figure S2. }
\par\end{centering}

Symmetries of the ground state (GS) and Split Kite (SK) structure
of a $\Sigma13(510)[001]$ grain boundary. (a) Grain boundary energy
as a function of atomic fraction measured of (510) plane. (b) We call
SK phase grain boundary structure that has glide symmetry, which is
not the ground state of $\Sigma13(510)[001]$ boundary at 0K. (c)
Ground state at 0.5 atomic fraction does not have glide symmetry.
The panels in (b) and (c) show two different views of the grain boundary
structure. Layer group symmetries exist in many of the generated grain
boundary configurations. \label{fig:symmtries}
\end{figure}

\begin{figure}
\begin{centering}
\includegraphics[width=0.6\textheight]{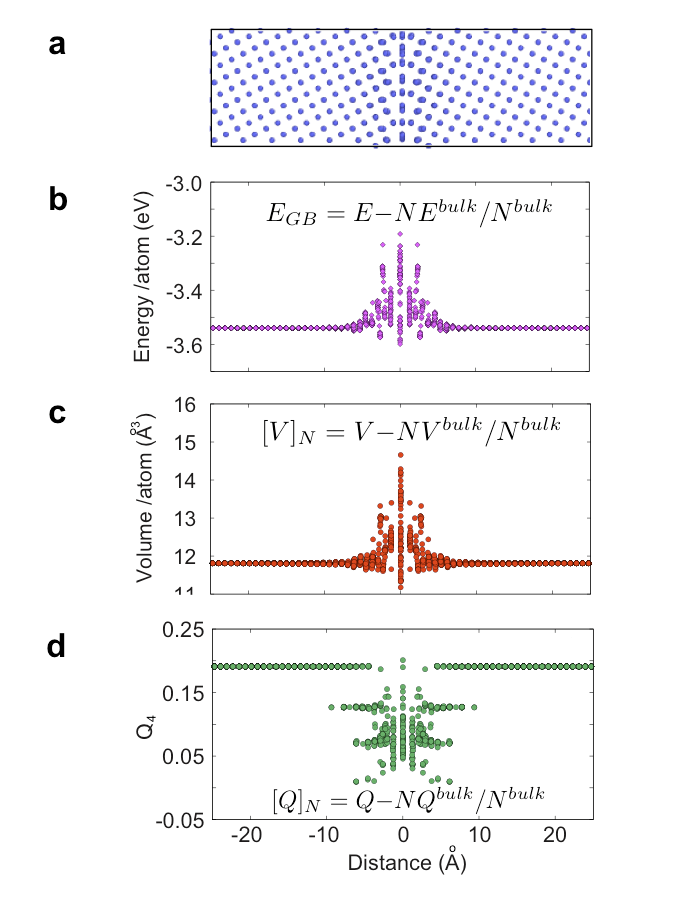} 
\par\end{centering}

\bigskip{}

\noindent \begin{centering}
\textbf{Supplementary Figure S3. }
\par\end{centering}

\protect

(a) A bicrystal with a grain boundary. (b) Energy per atom as a function
distance normal to the boundary. (c) Volume per atom as a fuction
distance normal to the boundary identified by Voronoi construction.
(d) Q4 order parameter calculated for each atom as a fuction distance
normal to the boundary. Properties in the boundary region are different
from the bulk. These data can be used to calculate excess properties
including $[E]_{N}$, $[V]_{N}$ and $[Q4]_{N}$ for each grain boundary
structure. \label{fig:prop_profs}
\end{figure}

\begin{figure}
\begin{centering}
\includegraphics[width=0.7\paperwidth]{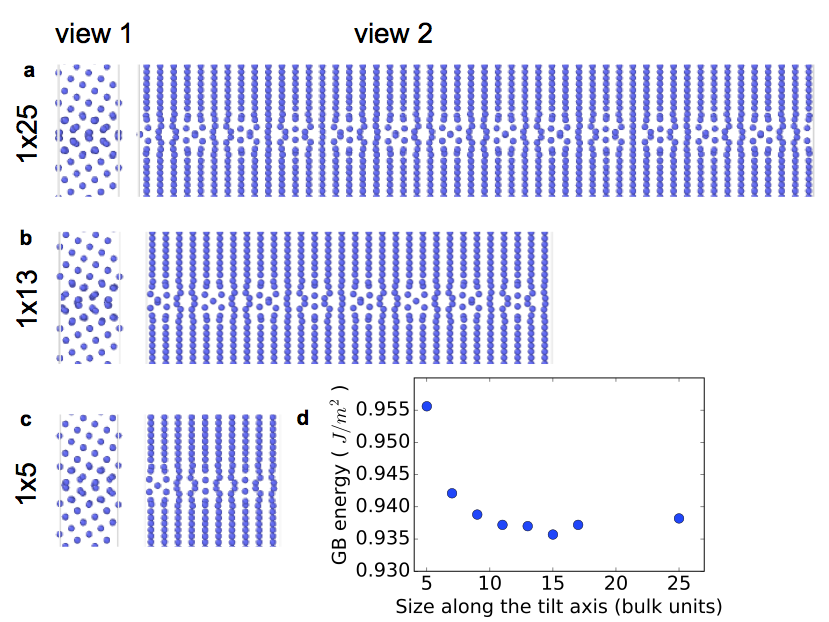} 
\par\end{centering}

\bigskip{}

\noindent \begin{centering}
\textbf{Supplementary Figure S4. }
\par\end{centering}

\protect

$1\times25$, $1\times13$ and $1\times5$ reconstructions of the
Split Kite phase of the $\Sigma5(210)[001]$ grain boundary. In the
left-hand side images (view 1) the {[}001{]} tilt axis is normal to
the plane of the screen. In the right-hand side images (view 2) the
{[}001{]} tilt axis is parallel to the plane of the screen. The different
reconstructions are composed of similar structural units and are nearly
indistinguishable to a eye in view 1. Grain boundary energy of different
reconstructions as a function of the dimension along the {[}001{]}
tilt axis. The energy plot demonstrates the need to explore grain
boundary areas much larger than the periodic unit of the bulk cell
to find the low-energy configurations. Distinct grain boundary configurations
with sizes 9 and larger have nearly degenerate energy. \label{fig:E_vs_size}
\end{figure}

\end{document}